\documentclass[11pt,letterpaper]{article}
\usepackage{amsmath,latexsym,amssymb,ifthen}
\usepackage{xspace}
\usepackage{graphicx,psfrag}
\usepackage{algorithm}
\usepackage{algpseudocode}
\newboolean{draft}
\setboolean{draft}{true}

\newcommand{\set}[1]{\left\{#1\right\}}

\newcommand{\proofstart}{{\noindent \bf Proof.\hspace{.2em}}}
\newcommand{\proofend}{\hspace*{\fill}\mbox{$\Box$}}

\def\ii_(#1,#2){i_{#1}^{#2}}

\def\qs{{\bf qs}}

\def\a{\alpha}
\def\b{\beta}

\def\g{\gamma}
\def\G{\Gamma}

\def\z{\zeta}

\def\th{\theta}

\def\l{\lambda}
\def\m{\mu}
\def\n{\nu}

\def\r{\rho}

\def\x{\xi}

\newcommand{\whp}{{\bf whp}\xspace}

\def\cE{\mathcal{E}}

\newcommand{\brac}[1]{\left( #1 \right)}

\newcommand{\expect}{\operatorname{\bf E}}
\def\E{\expect}
\renewcommand{\Pr}{\operatorname{\bf Pr}} 
\newcommand\bfrac[2]{\left(\frac{#1}{#2}\right)}

\newtheorem{theorem}{Theorem}
\newtheorem{lemma}[theorem]{Lemma}

\newtheorem{remthm}[theorem]{Remark}

\newtheorem{observation}[theorem]{Observation}
\newcounter{thmtemp}

\newcommand{\ignore}[1]{}
\newcommand{\nospace}[1]{}

\def\path{\operatorname{PATH}}

\def\BDTS{{\bf BDTS}}
\def\expdist{\ensuremath{\operatorname{Exp}}}
\def\EX{\expdist(1)}
\newcommand{\Bin}{\ensuremath{\operatorname{Bin}}}
\newcommand{\Be}{\ensuremath{\operatorname{Be}}}
\def\V{{\bf Var}}
\newcommand{\beq}[1]{\begin{equation}\label{#1}}
\def\eeq{\end{equation}}
\newcommand{\ZP}[1]{Z^{P}_{#1,n}}
\newcommand{\ZA}[1]{Z^{A}_{#1,n}}
\renewcommand{\Re}{\mathbb{R}}

\begin{document}

\title
{\vspace*{-.5in} Average case performance of heuristics for multi-dimensional assignment problems}

\author{Alan Frieze%
\thanks{Research supported by NSF grant DMS-6721878, 
Department of Mathematical Sciences,
Carnegie Mellon University, Pittsburgh PA15213, 
\hbox{e-mail}~{\small\texttt{alan@random.math.cmu.edu}}}
\and
Gregory B.~Sorkin\thanks{Department of Mathematical Sciences, 
IBM T.J.~Watson Research Center, Yorktown Heights NY 10598,
\hbox{e-mail}~{\small\texttt{sorkin@watson.ibm.com}}}}

\maketitle

\markboth{}{}

\begin{abstract}
We consider multi-dimensional assignment problems in a probabilistic setting. 
Our main results are: 
(i) A new efficient algorithm for the 3-dimensional planar problem,
based on enumerating and selecting from a set of ``alternating-path trees'';
(ii) A new efficient matching-based algorithm for 
the 3-dimensional axial problem.
\end{abstract}

\section{Introduction}
A (two-dimensional) assignment can be viewed as a set of pairs $P=\{(i_t,j_t),\,t=1,2,\ldots,n\}$ such that 
\begin{align} \label{matching}
\set{i_1,i_2,\ldots,i_n}=\set{j_1,j_2,\ldots,j_n}=[n] ,
\end{align}
i.e., each row appears once in $P$, as does each column
(and without loss of generality we may take
$(i_1,\ldots,i_n) = (1,\ldots,n)$).
Given an $n\times n$ matrix of costs $C=[C_{i,j}]$, 
the aim is to compute $P$ that minimises $C(P)=\sum_{(i,j)\in P}C_{i,j}$.
This is a well-studied problem from the worst-case
as well as the probabilistic point of view and it is solvable in polynomial time.

In the standard probabilistic model for the assignment problem
each entry $C_{i,j}$ is independently distributed
as the exponential random variable with mean one, viz., \EX. 
There are numerous results related to this 
model: If $Z_n$ is the minimum value of $C(P)$ then we have the remarkable result, conjectured by Parisi \cite{Parisi},
$$\E(Z_n)=\sum_{i=1}^n\frac{1}{i^2}.$$
This was proved by Linusson and W\"{a}stlund \cite{LiWa} and Nair, Prabhakar and Sharma \cite{NaPrSh}.
See W\"{a}stlund \cite{Wa1} for a remarkably short proof of this. Earlier work proving that 
$\lim_{n\to\infty}\E(Z_n)=\zeta(2)$ was done by Aldous \cite{Aldous1}, \cite{Aldous2}. 

In this paper we are concerned with the probabilistic analysis of multi-dimensional generalisations of this problem.
We consider two versions, Planar and Axial. Let us first consider the $d$-dimensional Planar model. 
Here we are given an $n \times \cdots \times n$ $d$-dimensional matrix (tensor) 
$C=[C_{i_1,i_2,\ldots,i_d}]$, i.e., a map $C:[n]^d\to\Re$. 
An assignment is a set 
of $n$ $d$-tuples $T=\{(\ii_(1,t),\ii_(2,t),\ldots,\ii_(d,t)),\,t=1,2,\ldots,n\}$ 
such that, in analogy with \eqref{matching},
for each dimension $r \in [d]$,
$\set{\ii_(r,1),\ii_(r,2),\ldots,\ii_(r,n)}=[n]$. 
Geometrically, for $r\in [d]$, let an {\em r-plane} be a set of
$d$-tuples of the form $[n]^{r-1}\times \set{x}\times [n]^{d-r}$ for some $x\in [n]$. In the case of $d=2$
a plane corresponds to a row or column of matrix $C$. An assignment $T$ 
is then a collection of $d$-tuples such that each of the
$dn$ planes contain exactly one $d$-tuple from $T$.

\begin{sloppypar}
The optimsation problem here is to compute an assignment $T$ that minimises
$C(T)=\sum_{(i_1,i_2,\ldots,i_d)\in T}C_{i_1,i_2,\ldots,i_d}$. This problem is NP-hard 
for $d\geq 3$ and the case of $d=3$ is one of the original problems listed in Karp \cite{Karp}. 
Some of its characteristics and applications are discussed in a recent book by Burkard, Dell'Amico and Martello \cite{BDM}.
Very little is known about the probabilistic behavior of the minimum $\ZP{d}$ of $C(T)$
for $d\geq 3$. Grundel, Oliveira, Pasiliao and
Pardalos \cite{GOPP} show that $\ZP{d}\to 0$ \whp\ in this case. 
At this point we can give some easy results on $\ZP{d}$ which we state as
\end{sloppypar}
\begin{theorem}\label{th1}
$$\Omega\bfrac{1}{n^{d-2}}\leq \ZP{d}\leq O\bfrac{\log n}{n^{d-2}}.$$
\end{theorem}
(All proofs are given in the body of the paper.)
The upper bound in this theorem is non-constructive, relying on recent work of Johansson, Kahn and Vu \cite{JKV}.
Our main result concerns a Bounded Depth Tree Search algorithm \BDTS($k$). Here $k$ is a parameter that refers
to the number of {\em levels} of search. It is unfortunate, but our approach only seems to give something interesting
for $d=3$. 
\begin{theorem}\label{th2}
Suppose that $1\leq k\leq \g\log_2\log n$ where $\g$ is any constant strictly less than 1/2. Then, \whp
\begin{description}
\item[(a)] Algorithm \BDTS($k$) runs in time $O(n^{2^{k+2}})$.
\item[(b)] The cost of the set of triples $T$ output by \BDTS($k$) satisfies 
$$C(T)=O(2^kn^{-1+\th_k}\log n)$$
where $\th_k=\frac{1}{2^{k+1}-1}$.
\end{description}
\end{theorem}
Note that for $k$ such as $\frac13 \log_2\log n$ this is a
``mildly exponential'' running time, $n^{O(\log n)}$,
yielding a solution which is an $O(n^{1/\log n})$ approximation to the
optimum.
After dealing with the Planar version, we will turn to the Axial version. 
Here we are again given an $n \times \cdots \times n$ 
$d$-dimensional matrix $C$. 
Geometrically, let a {\em line} be a 
set of $d$-tuples of the form $\set{i_1}\times \cdots\set{i_r}\times [n] \times \set{i_{r+2}}
\times \cdots\set{i_d}$ for some $r$ and $i_1,\ldots,i_r,i_{r+2}\ldots,i_d$. 
In the case of $d=2$
a line corresponds to a row or column of matrix $C$. An assignment $T$ 
is then a collection of $n^{d-1}$ $d$-tuples such that each of the
$dn^{d-1}$ lines contains exactly one $d$-tuple from $T$.

\begin{sloppypar}
The optimisation problem here is to compute an assignment $T$ that minimises
$C(T)=\sum_{(i_1,i_2,\ldots,i_d)\in T}C_{i_1,i_2,,\ldots,i_d}$. This problem is NP-hard 
for $d\geq 3$ and this was proved in Frieze \cite{Frieze}. 
We will prove
\end{sloppypar}
\begin{theorem}\label{th3}
The optimal solution value $\ZA{d}$ satisfies the following:
\begin{description}
\item[(a)] $\ZA{d}=\Omega(n^{d-2})$ \whp\ for $d\geq 3$.
\item[(b)] When $d=3$ there is a polynomial time algorithm that finds a solution with 
cost $Z$ where $Z=O(n\log n)$ \whp.
\end{description}
\end{theorem}
This leaves the following open questions:
\begin{description}
\item[P1] What are the growth rates of $\E[\ZP{d}]$ and $\E[\ZA{d}]$ for $d\geq 3$? 
\item[P2] Are there asymptotically optimal, polynomial time algorithms for solving these problems when $d\geq 3$.
\item[P3] Frieze \cite{FBLP} gave a bilinear programming formulation of the 3-dimensional planar problem. There
is a natural heuristic associated with this formulation (see appendix). 
What are its asymptotic properties? 
\end{description}
\subsection{Structure of the paper}
We deal with the Planar version in Section \ref{PV}. 
We start with the proof of Theorem \ref{th1} in Section \ref{PV1}. 
Our next task is to analyse \BDTS. 
We will analyse a three level version in Section \ref{PV2}. 
This provides intuition for the general case,
analysed in Section \ref{MG},
completing the proof of Theorem \ref{th2}. 
The Axial problem is considered in Section \ref{AV}.
The lower bound in Theorem \ref{th3} is proved in Section \ref{3GA1}
and the upper bound in Section \ref{3GA}.
\section{Multi-Dimensional Planar Version}\label{PV}
\subsection{Proof of Theorem \ref{th1}}\label{PV1}
Clearly 
$$\ZP{d}\geq \sum_{i_1=1}^n\min_{i_2,\ldots,i_d}C_{i_1,\ldots,i_d}.$$
Each term in the above sum is distributed as $\expdist(n^{d-1})$ and so has expectation $1/n^{d-1}$
and variance $1/n^{2d-2}$. The Chebyshev inequality implies that the sum is concentrated around the mean.

For the upper bound we use a recent result of Johansson, Kahn and Vu \cite{JKV}. This implies that \whp\
there is a solution that only uses $d$-tuples of weight at most $\frac{K\log n}{n^{d-1}}$. The upper bound
follows immediately. It should be noted that their proof is non-constructive.
\subsection{Two Level Version of BDTS}\label{PV2}
In this section we consider a two level version of the algorithm \BDTS. 
In this way we hope that to make it easier
to understand the general version that is described in Section \ref{MG}.
With reference to Theorem~\ref{th2},
the two-level version means taking $k=3$, $\th=\th_3=1/7$.

The heuristic has three phases:
\subsubsection{Greedy Phase}\label{G1}
The first phase is a simple greedy procedure. 
\\
{\bf Greedy Phase}
\begin{enumerate}
\item Let $n_1=n-n^{1-\th}$, $J=K=[n]$, and $T=\emptyset$.%
\footnote{We will often pretend that some expressions are integer. 
Formally, we should round up or down but it will not matter.}
\item For $i=1,\ldots,n_1$ do the following:
\begin{itemize}
\item Let $C_{i,j,k}=\min\set{C_{i,j',k'}:\;j'\in J,k'\in K}$;
\item Add $(i,j,k)$ to $T$ and remove $j$ from $J$ and $k$ from $K$.
\end{itemize}
\end{enumerate}
At the end of this procedure the triples in $T$ provide a partial assignment. Let
$$Z_1=\sum_{(i,j,k)\in T}C_{i,j,k}.$$
\begin{lemma}\label{lem1}
$$Z_1\leq \frac{2}{n^{1-\th}}\qquad \whp.$$
\end{lemma}
\proofstart
We observe that if $(i,j,k)\in I$ then $C_{i,j,k}$ is the minimum of $(n-i+1)^2$ independent 
copies of $\EX$ and is therefore distributed as $\expdist((n-i+1)^2)$. 
Furthermore, the random variables $C_{i,j,k}, (i,j,k)\in T$ are independent. 
Using the facts that an $\expdist{\l}$ random variable
has mean $1/\l$ and variance $1/\l^2$,
$$\E(Z_1)=\sum_{i=1}^{n_1}\frac{1}{(n-i+1)^2}\leq \int_{x=1}^{n_1+1}\frac{dx}{(n-x+1)^2}\leq \frac{1}{n^{1-\th}}.$$
Now 
$$\V(Z_1)=\sum_{i=1}^{n_1}\frac{1}{(n-i+1)^4}\leq \frac{3}{n^{3(1-\th)}}=o(\E(Z_1)^2)$$
and the lemma follows from the Chebyshev inequality.
\proofend
\subsubsection{Main Phase}\label{MP0}
The aim of this phase is to increase the size of the partial assignment defined by $T$ to $n-O(1)$.
Let $I=I(T)$ be the set of first coordinates assigned in $T$, i.e.,
$I=I(T) 
=\set{i:\;\exists j,k \text{ s.t. } (i,j,k)\in T}$. 
Relabeling if necessary,
without loss of generality we may assume that $I=[|T|]$.
This phase will be split into {\em rounds}. We choose a small constant $0<\a\ll 1$ and let $\b=1-\a$.
The aim of a round is to reduce the size of the 
set of unmatched first coordinates $X(T)=[n]\setminus I(T)$ by a
factor $\b$ while increasing 
the total cost of the matching only by an acceptably small amount. Thus we let 
$x_1=n-n_1$ and $x_t=\b^{t-1}x_1$ for $t\geq 2$. The aim of round $t$ is to reduce $|X(T)|$ from $x_t$ to $x_{t+1}$.
We continue this for $t_0=\log_{1/\b}(x_1/L)$ rounds where $L$ is a large positive constant. 
Thus at the end of 
the Main Phase, if successful, we will have a partial assignment of size at least $n-2L$.

So suppose now that we are at the start of a round and that $|X(T)|=x_t$. This is true for $t=1$. Next let 
$w_0=2n^{-12/7}\log n$ and
$$w_t=2n^{-6/7}x_t^{-8/7}\log^{1/7}n\qquad \text{for } t\geq 1.$$
At the start of each round we will {\em refresh} the array $C$ with independent exponentials, at some cost. 
By this we mean that we replace $C$ by a new array $C'$ where $C_{i,j,k}\leq C_{i,j,k}'+w_{t-1}$ and
the entries of $C'$ are i.i.d.\ \EX random variables.
More precisely, suppose that during the previous round we determined the precise values for all $C_{i,j,k}\leq w_{t-1}$
and left our state of knowledge for the other $C_{i,j,k}$ as being at least $w_{t-1}$. 
Then the memoryless property of exponentials means that 
\begin{align*}
C'_{i,j,k} = 
 \begin{cases}
  C_{i,j,k}-w_{t-1} & \text{when $C_{i,j,k} > w_{t-1}$}
  \\
  \text{fresh } X_{i,j,k} \sim \EX & \text{otherwise}
 \end{cases}
\end{align*}
has the claimed property.
Thus we can start a round with a fresh matrix of independent exponentials at the expense of adding another $w_{t-1}$ to
each cost. We note also that we can \whp\ carry out the Greedy Phase only looking at those $C_{i,j,k}$ of value less than
$w_0$.

Let $T_t$ denote the value of $T$ at the start of round $t$ and let $I_t=I(T_t),X_t=X(I_t)$.
In round $t$ we will add $A_t=[n-x_t+1,n-x_{t+1}]$ to $I_t$. By relabeling if necessary we will assume that
at the start of round $t$ we have $T=\set{(i,i,i):\;1\leq i\leq n-x_t}$. 
To add $i\in A_t$ to $I_t$ we find
distinct indices $j,k,p,q,r,s \in I_t$
(distinctness is not strictly necessary)
and replace 6 of the triples in $I_t$ by 7 new triples:
\begin{multline}\label{add}
+(i,j,k)-(j,j,j)-(k,k,k)+(j,p,q)+(k,r,s)-(p,p,p)-(q,q,q)-(r,r,r)-(s,s,s)+\\
(p,\x_1,\x_2)+(q,\x_3,\x_4)+(r,\x_5,\x_6)+(s,\x_7,\x_8),
\end{multline}
where $\x_1,\ldots,\x_8$ are distinct members of $X_t$,
and each of the triples added in \eqref{add} is required
to have (refreshed) cost at most $w_t$. 
Roughly, we are assigning a new 1-coordinate $i$,
this collides with previously used 2-coordinate $j$ and 3-coordinate $k$,
so the $(j,j,j)$ and $(k,k,k)$ elements 
are removed from the existing assignment,
1-coordinates $j$ and $k$ are re-added as $(j,p,q)$ and $(k,r,s)$
thus colliding with the previous assignment elements
$(p,p,p)$, $(q,q,q)$, $(r,r,r)$, and $(s,s,s)$,
and finally 1-coordinates $p$, $q$, $r$, $s$ 
are re-added as $(p,\x_1,\x_2)$ etc.,
where the $\x_i$ are elements \emph{not} previously assigned.
One may think of \eqref{add} as a binary tree version of 
an alternating-path construction;
we will control the cost despite the tree's expansion.

Putting $W_t=w_0+w_1+\cdots +w_t$ we see that if we can add one element
to $T$ at a cost of at most $w_t$ in refreshed costs, 
then in reality it costs us at most $W_t$;
step \eqref{add} increases the cost by $\leq 7 W_t$.
Success in a round means doing this $x_t-x_{t+1}$ times, in which
case the additional cost of the Main Phase will be at most 7 times
\begin{multline}\label{success}
\sum_{t=1}^{t_0}(x_t-x_{t+1})W_t\leq x_1(w_0+w_1)+\sum_{t=2}^{t_0}x_tw_t\\
\leq 3n^{-6/7}\log n+
2x_1^{-1/7}n^{-6/7}\log^{1/7}n\sum_{t=2}^{t_0}\b^{-t/7}\leq 4n^{-6/7}\log n.
\end{multline}
We must now show that \whp\ it is possible to add $x_t-x_{t+1}= \a x_t$ triples in round $t$ with a (refreshed)
cost of at most $7w_t$ per triple. For this we fix $t$ and drop the suffix $t$ from all quantities that use it.
We will treat refreshed costs as actual costs and drop the word ``refreshed''.

We start by estimating the number of choices for assigning $p$. Ignoring other indices, the number of choices is 
distributed as the binomial
$\Bin(\n,1-e^{-wx^2})=\Bin(\n,(1-o(1))wx^2)$ where $\n=n-x$. 
Here $1-e^{-wx^2}$ is the probability that for a given $p$, there exist $\xi_1,\xi_2$ such that 
$C_{p,\x_1,\x_2}\leq w$.
Note that 
$$wx^2=2(x/n)^{6/7}\log^{1/7}n=o(1) \text{ and that } wnx^2\gg \log n$$ 
and so the Chernoff bounds imply that,
\qs,\footnote{A sequence of events $\cE_n,n\geq 0$ are said to occur {\em
quite surely}, \qs, if $\Pr(\cE_n)=1-O(n^{-K})$
for any constant $K>0$.} we can choose a set $P$ of size exactly $wnx^2/2=o(n)$, such that 
for each $p\in P$ there is at least one
choice $\x_1,\x_2\in X$ such that the triple $(p,\x_1,\x_2)$ is {\em good},
i.e., $C_{p,\x_1,\x_2}\leq w$. 
Given this set of choices $P$ we 
find that the number of choices for $q\notin P$ is distributed as the binomial $\Bin(\n-|P|,1-e^{-wx^2})$ 
and we can once again
\qs\ choose a set $Q$, disjoint from $P$ such that $|Q|=wnx^2/2$ and each $q\in Q$ is in some good triple
$(q,\x_3,\x_4)$ where $\x_3,\x_4\in X$. Similarly, we can choose sets $R$, $S$
of choices for $r$, $s$, of size $wnx^2/2$,
such that $P,Q,R,S$ are pairwise disjoint.

\begin{observation}\label{obsa} 
Each $\x\in X$ is in $\Bin(x\n,1-e^{-w})$ good triples of the form 
$(p\in P,\x',\x'')$ and so
\qs\ it is in at most 
$$2wnx=\frac{4n^{1/7}\log^{1/7}n}{x^{1/7}}$$ 
such triples. 
\end{observation}
We now discuss our choices for $j$ and $k$. For a fixed $j$ there are $w^2n^2x^4/4$ pairs in $P\times Q$ and
each has a probability $1-e^{-w}$ of forming a good triple $(j,p,q)$. Let $j$ be {\em useful} if there is such a pair
and {\em useless} otherwise. Then
$$\Pr(j \text{ is useless})\leq \exp\set{-\frac{w^3n^2x^4}{4}}\leq 1-\frac{w^3n^2x^4}{5}.$$
It follows that the number of useful $j\notin Y=P\cup Q\cup R\cup S$ 
dominates $\Bin(n-o(n),w^3n^2x^4/5)$ and so \qs\ we can choose a set 
$J$ of useful $j\notin Y$ of size 
$$\frac{w^3n^3x^4}{6}=\frac{4n^{3/7}x^{4/7}\log^{3/7}n }{3}=o(n).$$ 
We can by a similar argument choose a set $K$ of useful $k$ 
of this size disjoint from $J$ and $Y$.

\begin{observation}\label{obsb}
A fixed $p$ is in at most $\Bin(wn^2x^2/2,1-e^{-w})$ good triples $(j,p,q)$ where $(j,q)\in 
J\times Q$ and so \qs\ every $p$ is in at most $w^2n^2x^2$ such triples. 
\end{observation}
Suppose then that in the middle of a round we have added $y<\a x$ triples to $T$. The number of
$\x\in X$ that can be used in a good triple $(p,\x,\eta)$ will have been reduced by $y$. The number
of $\eta$ will have been reduced by the same amount. It follows from Observation 
\ref{obsa} that the number of choices for $p$ will have been 
reduced by at most $2\a x\times 2wnx$. By Observation \ref{obsb} this reduces the number of choices for $j$ by at most
$2\a x\times 2wnx\times w^2n^2x^2+7\a x\ll |J|=w^3n^3x^4/6$. The additional term $+7\a x$ accounts for
the choices we lost because they have previously been used in this round.
So our next $i$ will get a choice of at least
$\Bin((w^3n^3x^4/7)^2,1-e^{-w})$ choices for a good triple $(i,j,k)$. So the expected number of choices is at least
$w^7n^6x^8/49=(2^7/49)\log n$ and then the probability there is no choice is $o(n^{-1})$.
This is sufficient to ensure that \whp\ there is always at least 
one choice for every $i$.

\subsubsection{Final Phase}
\newcommand{\xx}{{n}}
We now have to add only $O(1)$ indices to $I$. 
At this point there is a problem with the bottom-up approach of the 
previous phase if $x<8$, clearest in the case $x=1$, say the single element $\xx$,
when each of
$\x_1,\ldots,\x_8$ would have to be $\xx$,
leading to an illegal assignment.
Thus instead we will work top down.
The details of this will cause more conditioning of the matrix, 
and therefore
we refresh $C$ 
after each increase in $I$,
at an extra cost of $w=Kn^{-6/7}\log^{1/7}n$.
So, if successful, the cost of this round is 
$O(W_{t_0}+w)=O(n^{-6/7}\log^{1/7}n)$. 

Let us now replace the notation of \eqref{add} by 
\begin{multline}\label{add1}
+(i,j,k)-(j_1,j,j_3)-(k_1,k_2,k)+(j_1,p,q)+(k_1,r,s)-(p_1,p,p_3)-(q_1,q_2,q)-(r_1,r,r_3)-(s_1,s_2,s)+\\
(p_1,i_2,p_3)+(q_1,q_2,j_3)+(r_1,s_2,i_3)+(s_1,k_2,r_3),
\end{multline}
where any subtracted triple such as $(j_1,j,j_3)$
denotes a previous match 
(we are no longer assuming the convention that such a triple
would be $(j,j,j)$),
and where $i_2,i_3$ are unused 2- and 3-coordinates respectively.

Fix $j$ (and thus its previously matched companion indices $j_1, j_3$)
and let $Z_j$ be the number of choices for $p,q$ 
(with their previously matched companion indices $p_1,p_3,q_1,q_2$)
such that 
$C(j_1,p,q),C(p_1,i_2,p_3),C(q_1,q_2,j_3)\leq w$. 
This has the distribution $B_1(B_2(n,w)B_3(n,w),w)$ where $B_1,B_2,B_3$ denote independent binomials, 
with $B_2$ counting the good choices for $p$, $B_3$ those for $q$,
and $B_1$ those for $j$ using these $p$ and $q$ possibilities.
Using Chernoff bounds on the binomials $B_2,B_3$ we see that \whp\ $Z_j$ dominates $B(n^2w^2/2,w)$ which dominates 
$\Be(n^2w^3/3)$, the Bernoulli random variable that is 1
with probability $n^2w^3/3$ and 0 otherwise. 
The same holds for index $k$ and \eqref{add1} has been constructed so that choices for $j,k$ are independent. So, the number of choices
for $j,k$ dominates $\Bin(n^2,w(n^2w^3/3)^2)$ which has expectation $\Omega(\log n)$ and so is non-zero \whp.

This completes the analysis of \BDTS\ when there are two levels.
\subsection{General 3-Dimensional Version}\label{MG}
We follow the same three phase strategy. $k$ is a positive integer, $2\leq k\leq \g\log\log n$.

\subsubsection{Greedy Phase}
This is much as before. 
Proceed as in Section \ref{G1}
but taking $\th=\th_k$ (recall $\th$'s definition from Theorem~\ref{th2}) 
and defining $n_1$ accordingly.
Lemma \ref{lem1} continues to hold.

\subsubsection{Main Phase}\label{MP1}
Let 
$$\a=2^{-2k-2}\brac{1-\sqrt{2/3}}$$
and let $\b,t_0$ and $x_t,t=1,\ldots,t_0$ be defined as in Section \ref{MP0}.
Let $I_t,X_t,A_t$ have the same meaning as well.
Now let $w_0=2n^{-2(1-\th_k)}\log n$ and
$$w_t=2x_t^{-1-\th_k}n^{\th_k-1}\log^{\th_k} n\qquad \text{for } t\geq 1$$
and 
$$W_t=w_0+w_1+\cdots +w_t=O\bfrac{\log^{\th_k}n}{n^{1-\th_k}}.$$
The aim of round $t$ is once again to add $x_t-x_{t+1}$ new indices to $I_t$ using triples with (refreshed) cost at most
$w_t$. We will assume that at the start of round $t$ we have $T=\set{(i,i,i):\;1\leq i\leq n-x_t}$.
In analogy with \eqref{add},
to add $i\in A_t$ to $I_t$ we will add $2^{k+1}-1$ triples to $T$
and remove $2^{k+1}-2$ triples,
in which case the additional cost of the Main Phase will be at most $2^{k+1}-1$ times
\begin{multline}\label{successg}
\sum_{t=1}^{t_0}(x_t-x_{t+1})W_t\leq x_1(w_0+w_1)+\sum_{t=2}^{t_0}x_tw_t\\
\leq 3n^{\th_k-1}\log n+
2x_1^{-\th_k}n^{\th_k-1}\log^{\th_k}n\sum_{t=2}^{t_0}\b^{-\th_kt}
\leq 4n^{\th_k-1}\log n.
\end{multline}
The notation used in \eqref{add} is obviously insufficient. We imagine a rooted tree $\G$ of triples. The root
will be $\r=(i_0,j_0,k_0)$ where $i_0$ is the index to be added to $I_t$. The root is at level zero. The triples
at odd levels are to be deleted from $T$ and the vertices at even levels are to be added to $T$. Every
triple at an odd level $2l-1$ will therefore have the form $(p,p,p)$ where $p\in I_t$. This triple will have one child
$(p,a,b)$ which will replace the parent triple in 1-plane $p$. If $l<k$ then $a,b\in I_t$ and if $l=k$ then $a,b\in X_t$.
A triple $u=(p,a,b)$ at an even level will have two children. By construction, $u$ will be the unique triple in 1-plane
$p$, but now we will have two triples in 2-plane $a$ and 3-plane $b$. Thus the children of $u$ are $(a,a,a)$ and $(b,b,b)$.
This defines a tree corresponding to adding $2^{k+1}-1$ and removing $2^{k+1}-2$ triples from $T$. 
We ensure that if $u=(p,a,b)$ is a triple at an even level, then $p,a,b$ do not appear anywhere else in the tree, except at
the child of $u$ as previously described. We do this so that additions in one
part of the tree do not clash with additions in another part and then the additions and deletions give rise to a partial
assignment. We also insist that if $u=(p,a,b)$ is a triple at an even level then $C_{p,a,b}\leq w$. We call such a tree
{\em feasible}. We considered each level of $\G$ to be ordered so it makes sense to talk of the $r$th vertex of level $2l$ 
where $1\leq r\leq 2^l$.

We now have to show that \whp\ there is always at least one such tree $\G$ for each $i\in A_t$. 
We take the same {\em bottom-up} approach that we did in Section \ref{PV2}.
We fix $t$ and drop the suffix $t$ from all quantities that use it.
We start by estimating the number of choices for a $p$ that can be in a triple $(p,x,y)$ at level $2k$. 
Ignoring other indices, the number of choices is again
distributed as the binomial
$\Bin(\n,1-e^{-wx^2})=\Bin(\n,(1-o(1))wx^2)$ where $\n=n-x=n-o(n)$. Note that $wx^2=K(x/n)^{1-\th_k}\log^{\th_k}n=o(1)$ and that
$wnx^2=\tilde{\Omega}(n^{\th_k})\gg\log n$. (Here our notation $f(n)\gg g(n)$ means that $f(n)/g(n)\to\infty$ with $n$). So 
the Chernoff bounds imply that 
\qs\ we can choose a set $P$ of size exactly $wnx^2/2=o(n)$, such that 
for each $p\in P$ there is at least one
choice $\x_1,\x_2$ such that the triple $(p,\x_1,\x_2)$ is {\em good}, i.e., $C_{p,\x_1,\x_2}\leq w$. 
We will in fact be able to choose $2^k$ disjoint sets $P_{l,k},1\leq l\leq 2^k$ since replacing $\n$ by
$\n-2^kwnx^2/2$ will not significantly change the above calculations. (Here $2^kwnx^2=O(n^{1-\th_k+\th_k^2}\log^{\th_k+\g}n)=o(n)$).

\begin{observation}\label{obs1} 
Each $\x\in X$ is in $\Bin(x\n,1-e^{-w})$ good triples of the form $(p\in P_{l,k},\x,.)$ and so
\qs\ it is in at most $2wnx$ such triples. (Here $wnx=2\bfrac{n\log n}{x}^{\th_k}\gg\log n$).
\end{observation}
Let 
\beq{recur}
\n_0=wnx^2/2 \text{ and } \n_{l+1}=wn\n_l^2/2 \text{ for } 0\leq l< k.
\eeq
 The solution to this recurrence is
$$\n_l=\bfrac{wn}{2}^{2^{l+1}-1}x^{2^{l+1}}=(n\log n)^{(2^{l+1}-1)\th_k}x^{(2^{k+1}-2^{l+1})\th_k}.$$
Observe that $\n_l$ increases with $l$.
Note also that if $l\leq k-2$ then 
\begin{align}
&w\n_l^2\leq w\n_{k-2}^2=2\bfrac{x}{n}^{2^k\th_k}\log^{(2^k-1)\th_k}n=o(1),\label{use}\\
&wn\n_l\geq wn\n_0=\frac{w^2n^2x^2}{2}=2\bfrac{n\log n}{x}^{2\th_k}\gg\log n.\label{usex}
\end{align}
We now have the basis for an inductive claim that \qs\ if $l\leq k-1$ and $u=(p,a,b)$ is a triple at an even level $2(k-l)$ then there 
are at least $\n_l$ choices for $p$ such that there exists a triple $u=(p,a,b)$ with $C_u\leq w$ and a feasible tree
$\G_u$ with $u$ as root and depth $2l+1$. Our analysis above has proved the base case of $l=0$. 
Imagine now that we are filling in the possibilities for the $r$th triple $(p,a,b)$ at level $k-l$.
We fill in these possibilities level by level starting at level $2k$.
Imagine also that we have identified
$\n_{l-1}$ choices for each of $a,b$. 
This can be an inductive assumption, so for example $a$ 
will have to be a possible selection
for the first component of the $(2r-1)$st triple at level $2(k-(l-1))$.

For a fixed $p$, conditional
on our having selected exactly $\n_{l-1}$ choices $A,B$ for $a,b$,
let $p$ be {\em useful} if there is a pair $(a,b)\in A\times B$ 
with $C_{p,a,b}\leq w$
and {\em useless} otherwise. Then, using \eqref{use},
$$\Pr(p \text{ is useless})\leq \exp\set{-w\n_{l-1}^2}\leq 1-\frac{2w\n_{l-1}^2}{3}.$$
It follows that the number of useful $p$ that have not been previously selected
dominates $\Bin(n-o(n),2w\n_{l-1}^2/3)$. Here $o(n)=\sum_{s\leq l}2^{k-s}w\n_s^2$ bound the number of {\em forbidden} $p$'s. 
It follows that \qs\ we can choose a set 
of useful $p$'s of size $w\n_{l-1}^2/2=o(n)$. We can do this so that each node of $\G$ gets distinct choices. 

\begin{observation}\label{obs2}
A fixed $a$ is in at most $\Bin(n\n_{l-1}/2,1-e^{-w})$ good triples $(p,a,b)$ feasible for level $2(k-l)$
and so \qs\ every $a$ is in at most $wn\n_{l-1}$ such triples, see \eqref{usex}. 
\end{observation}
This completes our induction. We now apply the above to show that round $t$ succeeds \whp.

Suppose that in the middle of a round we have added $y<\a x$ triples to $T$. The number of
$\x\in X$ that can be used in a good triple $(p,\x,\eta)$ at level $2k$ will have been reduced by $y$. 
Thus the number of choices for $p$ in any triple in this level will have been 
reduced by at most $2^k\times 2\times \a x\times 2wnx$, see Observation \ref{obs1}. This reduces the number of choices for $p$ 
in a triple at level $2(k-1)$ by at most
$2^{k+2}\a wnx^2\times wn\n_0=2^{k+3}\a wn\n_0^2$, see Observation \ref{obs2}. 
So let $\m_l$ denote the number of choices for $p$ in triples $p(,.,.)$ at level $2(k-l)$ that are forbidden by choices
further down the tree. We have just argued that $\m_1\leq 2^{k+3}\a wn\n_0^2$. In general we can use Observation
\ref{obs2} to conservatively argue that 
$$\m_l\leq wn\n_{l-1}(\m_{l-1}+2^{k+1}\a x).$$
It follows that for $l\geq 2$ we have 
\begin{align*}
\frac{\m_l}{\n_l}&\leq 2\frac{\m_{l-1}}{\n_{l-1}}+\frac{2^{k+2}\a x}{\n_{l-1}}
\leq 2\frac{\m_{l-1}}{\n_{l-1}}+2^{k+2}\a\bfrac{x}{n\log n}^{(2^l-1)\th_k}
\leq 2\frac{\m_{l-1}}{\n_{l-1}}+2^{k+2}\a\bfrac{x}{n\log n}^{\th_k}.
\end{align*}
It follows that
$$\frac{\m_{k-1}}{\n_{k-1}}\leq 2^{k-2}\frac{\m_1}{\n_1}+2^{2k+1}\a\bfrac{x}{n\log n}^{\th_k}
\leq 2^{2k+2}\a.$$

We see that at the root there will still be at least $(1-2^{2k+2}\a)\n_{k-1}$ choices for $j_0,k_0$.
So $i_0$ will get a choice of at least
$\Bin((1-2^{2k+2}\a)^2\n_{k-1}^2,1-e^{-w})$ choices for a good triple $(i_0,j_0,k_0)$. 
So the expected number of choices is at least
$2w\n_{k-1}^2/3$, our choice of $\a$ implies this. Now
$w\n_{k-1}^2=2\log n$
and this is sufficient to ensure that \whp\ there is always at least 
one choice for every $i_0$.
\subsubsection{Final Phase}
We can execute the Main Phase so long as $x\geq 2^k$. Now assume that $1\leq x<2^k$.
We now have to add only $O(1)$ indices to $I$. This time we refresh $C$ an $O(2^k)$ number of times at an extra cost of
 $w=\displaystyle \frac{\log^{\th_k}n}{n^{1-\th_k}}$
each time we add an index. 
So, if successful, the cost of this round is $O(W_{t_0}+w)=O\bfrac{\log^{\th_k}n}{n^{1-\th_k}}$. 

We first make an inductive assumption: We have a partial assignment $I$ where $|I|\leq n-2$. 
(The reader might think that we should assume $|I|\leq n-1$, but here we use the induction hypothesis
after one more index has temporarily been deleted from $I$, prior to a replacement). Assume that 
the matrix $C$ is unconditioned and $i\notin I$:
Then we can in $O(n^{2\ell})$ time 
\whp\ find a set $P$ of size $\n_{\ell-1}$ (with $x=1$ in definition \eqref{recur}) and a collection $Q_p,p\in P$
of sets of size $\n_{\ell-1}$
such that for each $(p,q\in P_p)$ 
there is an assignment $P'$ with $(i,p,q)\in P'$ and $I(P')\supsetneqq I(P)$ and $C(P')=C(P)+C(i,p,q)+O(w)$.
This is true for $\ell=1$ since we can make the changes
$$+(i,p,q)-(p_1,p,p_3)-(q_1,q_2,q)+(p_1,i_2,p_3)+(q_1,q_2,i_3)$$
where $i_2,i_3$ are unused 2- and 3-coordinates respectively. The number of
choices for $p,q$ are independent $\Bin(n,w)$.

For the inductive step, we first refresh the matrix $C$. Then for each $p\in [n]$ we let $I'=I-\set{p}$ and apply the 
induction hypothesis to generate $\n_{\ell-2}^2$ choices of assignment that add back $p_1$ to $I'$. We find that
\whp\ at least $w\n_{\ell-2}^2/2=\n_{\ell-1}$ of these have $C(p_1,.,.)\leq w$. Let this set be $P$.
Now refresh $C$ again and apply the 
same argument for each $p\in P$ to generate choices $Q_p$ for $p$. This completes the induction.

Now let $\ell=k$ and refresh $C$ one more time. Let $P,Q_p,p\in P$ be the sets of size $\n_{k-1}$ promised by the above argument.
We have $\Bin(\n_{k-1}^2,w)$ choices of $j,k$ which can be
used to add $i\notin I$ to $I$ at a cost of $O(w)$. In expectation this is $2\log n$ and so we succeed \whp.
\ignore{
Following a similar analysis as for the Main Phase we have \whp\ at least $\n_{k-1}$ choices for each of $j_0,k_0$ for each
index added since $x\geq 1$ until the end. (The analysis is simplified by the fact that we refresh $C$ for each index,
obviating the need for using Observations \ref{obs1} and \ref{obs2}).
Thus for each $i_0$ there will be at least $\Bin(\n_{k-1}^2,1-e^{-w})$
choices for $j_0,k_0$. 
Here we use recurrence \eqref{recur} with $x=1$, yielding 
$$\n_{k-1}=\bfrac{wn}{2}^{2^k-1}=\bfrac{n\log n}{2}^{\th_k}.$$
In expectation, $\Bin(\n_{k-1}^2,1-e^{-w})\to\infty$ and so we succeed \whp.}

For the execution time of the algorithm we simply bound the number of possible trees $\G$.
This completes the proof of Theorem \ref{th2}.

\section{Multi-Dimensional Axial Version}\label{AV}
We turn to the proof of Theorem \ref{th3}. 
\subsection{Lower bound}\label{3GA1}
It is clear that $\ZA{d}\geq Z_1+Z_2+\cdots+Z_{n^{d-2}}$ where $Z_i$ is the
minimum cost of the 2-dimensional assignment with
cost matrix $A_{j,k}=C_{i_1,\ldots,i_{d-2},j,k}$. We know that $Z_j\geq (1-o(1))\z(2)$ \whp\ and the $Z_i$'s are independent. It follows that
\whp\ $\ZA{3}\geq (1-o(1))n^{d-2}\z(2)>3n^{d-2}/2$. 
\subsection{Upper bound for $d=3$}\label{3GA}
For the upper bound we need a result of Dyer, Frieze and McDiarmid \cite{DFM}. We will not state it in full generality,
instead we will tailor its statement to precisely what is needed.
Suppose that we have a linear program 
$$P:\qquad \text{Minimize $c^Tx$ subject to $Ax=b$, $x\geq 0$.}$$
Here $A$ is an $m\times n$ matrix and the cost vector $c=(c_1,c_2,\ldots,c_n)$ is a sequence of independent copies of $\EX$.
Let $Z_P$ denote the minimum of this linear program. Note that
$Z_P$
is a random variable. Next let $y$ be {\em any} feasible solution to $P$.
\begin{theorem}[\cite{DFM}]\label{DFM}
\beq{zz}
\E(Z_P)\leq m\max_{j=1,2,\ldots,n}y_j.
\eeq
Furthermore, $Z_P$ is at most $1+o(1)$ times the RHS of \eqref{zz}, \whp.
\end{theorem}
Now consider the following greedy-type algorithm. 
We find a minimum 2-dimensional assignment for
1-plane $i=1$, we then find a minimum assignment for 1-plane $i=2$, consistent
with choice for 1-plane $i=1$,
and so on:\\
{\bf Greedy}
\begin{enumerate}
\item For $i=1,\ldots,n$ do the following:
\begin{itemize}
\item Let $G=K_{n,n}\setminus (M_1\cup M_2\cup \cdots M_{i-1})$;
\item If $(j,k)\in E(G)$ let $A_{j,k}=C_{i,j,k}$.
\item Let $M_i$ be a minimum cost matching of $G$ using edge weights $A$.
\end{itemize}
\end{enumerate}
The output, $M_1,M_2,\ldots,M_n$ defines a set of triples $T=\set{(i,j,k):\;(j,k)\in M_i}$.
We claim that if $Z_i=A(M_i)$ then 
\beq{ax}
\E(Z_i)\leq \frac{2n}{n-i+1}.
\eeq
For this we apply Theorem \ref{DFM} to the linear program
\begin{align*}
\text{Minimise} \sum_{(j,k)\in E(G)}A_{j,k}x_{j,k} \quad & \text{ subject to} \\
\sum_{k:\;(j,k)\in E(G)}x_{j,k}=1, \qquad & j=1,2,\ldots,n\\
\sum_{j:\;(j,k)\in E(G)}x_{j,k}=1, \qquad & k=1,2,\ldots,n\\
x_{j,k}\geq 0, \qquad & j,k=1,2,\ldots,n .
\end{align*}
We note that there are $2n$ constraints and that $x_{j,k}=1/(n-i+1)$ is a
feasible solution. 
With Theorem~\ref{DFM}, this implies \eqref{ax} and the upper bound in Theorem \ref{th3} for the 
case $d=3$.

\appendix
\section{Bilinear Programming Formulation}
Frieze \cite{FBLP} re-formulated the 3-dimensional planar problem as
$$\text{Minimise} \sum_{i,j,k=1}^{n}C_{i,j,k}y_{i,j}z_{i,k} 
\text{ subject to } x,y\in P_A$$
where $P_A$ is the bipartite matching polyhedron $\sum_{i=1}^nx_{i,j}=1=\sum_{j=1}^nx_{i,j}$, for all $1\leq i,j\leq n$.

Now denote the objective above by $C(y,z)$. The following heuristic was used successfully in a practical situation \cite{FY}:
\begin{enumerate}
\item Choose $y_0,z_0$ arbitrarily; $Z_0=C(y_0,z_0)$; $i=0$.
\item Repeat until $Z_{i+1}=Z_i$.
\begin{itemize}
 \item Let $y_{i+1}$ maximise $C(y,z_i)$.
 \item Let $z_{i+1}$ maximise $C(y_{i+1},z)$.
 \item $Z_{i+1}=C(y_{i+1},z_{i+1})$.
 \item $i=i+1$.
\end{itemize}

\end{enumerate}


\begin{thebibliography}{99}
\bibitem{Aldous1} D. Aldous, Asymptotics in the random assignment problem, 
{\em Probability Theory and Related Fields} 93 (1992) 507-534.
\bibitem{Aldous2} D. Aldous, The $\z(2)$ limit in the random assignment problem, 
{\em Random Structures and Algorithms} 18 (2001) 381-418.
\bibitem{BDM} R. Burkard, M. Dell'Amico and S. Martello, Assignment Problems, SIAM Publications 2009.
\bibitem{DFM} M.E. Dyer, A.M. Frieze and C. McDiarmid, Linear programs with random costs,
{\em Mathematical Programming} 35 (1986) 3-16.
\bibitem{FBLP} A. M. Frieze, A bilinear programming formulation of the 3-dimensional assignment problem,
{\em Mathematical Programming} 7 (1974) 376-379.
\bibitem{Frieze} A. M. Frieze, Complexity of a 3-dimensional assignment problem,
{\em European Journal of Operational Research} 13 (1983) 161-164.
\bibitem{FY}  A. M. Frieze and J. Yadegar, An algorithm for solving 3-dimensional assignment problems with application to scheduling 
a teaching practice, {\em Journal of the Operational Research Society} 32 (1981) 989-995.
\bibitem{GOPP} D. Grundel, C. Oliveira, E. Pasiliao and E. Pardalos,
Asymptotic Results for Random Multidimensional Assignment problems,
{\em Computational Optimization and Applications} 30 (2005) 275-293.
\bibitem{JKV} A. Johansson, J. Kahn and V. Vu, Factors in random graphs,
{\em Random Structures and Algorithms} 33 (2008) 1-28.
\bibitem{Karp} R.M. Karp, Reducibility among combinatorial problems,
in R.E. Miller and J.W. Thatcher (Eds), Complexity of Computer Computations, Plenum Press, New York (1972) 85-103.
\bibitem{LiWa}  S. Linusson and J. W\"astlund, A proof of Parisi's conjecture on the random assignment problem,
{\em Probability Theory and Related Fields} 128 (2004) 419-440.
\bibitem{NaPrSh} C. Nair, B. Prabhakar and M. Sharma, Proofs of the Parisi and Coppersmith-Sorkin
conjectures for the finite random assignment problem, {\em Proceedings of IEEE FOCS} (2003) 168-178.
\bibitem{Parisi} G. Parisi, A conjecture on random bipartite matching, arXiv:cond-mat/9801176, 1998.
\bibitem{Wa1} J. W\"astlund, A simple proof of the Parisi and Coppersmith-Sorkin formulas for the 
random assignment problem, 
{\em Linköping Studies in Mathematics} 6 (2005). 
\end{thebibliography}
\end{document}